\documentclass[conference]{IEEEtran}
\usepackage{cite}
\usepackage{amsmath,amssymb,amsfonts}
\usepackage{algorithmic}
\usepackage{graphicx}
\usepackage{textcomp}
\usepackage{xcolor}
\usepackage{balance}
\def\BibTeX{{\rm B\kern-.05em{\sc i\kern-.025em b}\kern-.08em
    T\kern-.1667em\lower.7ex\hbox{E}\kern-.125emX}}

\usepackage{caption}
\usepackage{subcaption}
\usepackage{url}
\usepackage{enumitem}
\usepackage{booktabs}
\usepackage{array}
\usepackage{quoting}
\usepackage{hyperref}
\newcolumntype{L}[1]{>{\raggedright\let\newline\\\arraybackslash\hspace{0pt}}m{#1}}
\newcolumntype{C}[1]{>{\centering\let\newline\\\arraybackslash\hspace{0pt}}m{#1}}
\newcolumntype{R}[1]{>{\raggedleft\let\newline\\\arraybackslash\hspace{0pt}}m{#1}}

\usepackage{amssymb}
\usepackage{amsfonts}

\makeatletter
\def\ps@IEEEtitlepagestyle{
	\def\@oddfoot{\mycopyrightnotice}
	\def\@evenfoot{}
}
\def\mycopyrightnotice{
	{\footnotesize
		\begin{minipage}{\textwidth}
			\centering
			\textcopyright~2023 IEEE.  Personal use of this material is permitted.  Permission from IEEE must be obtained for all other uses, in any current or future media, including reprinting/republishing this material for advertising or promotional purposes, creating new collective works, for resale or redistribution to servers or lists, or reuse of any copyrighted component of this work in other works.
		\end{minipage}
	}
}

\usepackage[many]{tcolorbox}   
\newtcolorbox{boxA}{
    colback = white,
    boxrule = 1pt,
    colframe = black, 
    boxsep=1pt, 
    left=2pt,
    right=2pt,
    top=2pt,
    bottom=2pt
}
\begin{document}

\title{Runtime Verification of Self-Adaptive Systems with Changing Requirements}

\author{
	\IEEEauthorblockN{
            Marc Carwehl\IEEEauthorrefmark{1},
		Thomas Vogel\IEEEauthorrefmark{1},
		Genaína Nunes Rodrigues\IEEEauthorrefmark{2},
		and
		Lars Grunske\IEEEauthorrefmark{1}
	}
	\IEEEauthorblockA{\IEEEauthorrefmark{1}\textit{Department of Computer Science}, \textit{Humboldt-Universit\"at zu Berlin}, Berlin, Germany\\
		Email: carwehl@informatik.hu-berlin.de, thomas.vogel@informatik.hu-berlin.de, grunske@informatik.hu-berlin.de}
	\IEEEauthorblockA{\IEEEauthorrefmark{2}\textit{Department of Computer Science, University of Bras\'ilia, Brazil}\\
		Email: genaina@unb.br}
}

\maketitle

\begin{abstract}

To accurately make adaptation decisions, a self-adaptive system needs precise means to analyze itself at runtime. To this end, runtime verification can be used in the feedback loop to check that the managed system satisfies its requirements formalized as temporal-logic properties. These requirements, however, may change due to system evolution or uncertainty in the environment, managed system, and requirements themselves. Thus, the properties under investigation by the runtime verification have to be dynamically adapted to represent the changing requirements while preserving the knowledge about requirements satisfaction gathered thus far, all with minimal latency. To address this need, we present a runtime verification approach for self-adaptive systems with changing requirements. Our approach uses property specification patterns to automatically obtain automata with precise semantics that are the basis for runtime verification. The automata can be safely adapted during runtime verification while preserving intermediate verification results to seamlessly reflect requirement changes and enable continuous verification. We evaluate our approach on an Arduino prototype of the Body Sensor Network and the Timescales benchmark. Results show that our approach is over five times faster than the typical approach of redeploying and restarting runtime monitors to reflect requirements changes, while improving the system's trustworthiness by avoiding interruptions of verification.

\end{abstract}


\section{Introduction}

In a self-adaptive system (SAS), a managing system dynamically controls a managed system so that the managed system satisfies its requirements despite uncertainty concerning the environment, system, and requirements themselves~\cite{weyns2020introduction}. For this purpose, the managing system typically implements a MAPE-K feedback loop~\cite{kephart2003vision}. The monitoring and analysis steps observe the managed system and check whether the system satisfies its requirements. If not, the planning step creates an adaptation plan that the execution step will enact on the managed system to perform self-adaptation. In this context, one means for the analysis step to detect violations of requirements by the managed system is \textit{runtime verification}. 

We consider runtime verification methods that check the current execution of a system represented by a trace against the requirements formalized as \textit{properties} in a temporal logic such as Linear Temporal Logic (LTL) and Metric Temporal Logic (MTL)~\cite{bauer2011runtime,bartocci2018introduction}. 
Such a runtime verification has been already applied to self-adaptive systems to detect violations of properties~\cite{CailliauL19:TAAS,goldsby2007amoeba}. A further line of research focuses on stochastic behavior in self-adaptive systems, for which the violations of properties expressed in a probabilistic temporal logic can be detected by quantitative verification~\cite{Calinescu2012quantitative, calinescu2017using, filieri2011run, Solano+2019}. 

For SAS, we claim that any runtime verification performed by the managing system has to take into account uncertainties of the environment, managed system, and requirements. Particularly, any uncertainty may directly or indirectly cause changes of the requirements. Consequently, the properties under investigation by the runtime verification have to be dynamically adapted to represent the changing requirements. 
For example, in the healthcare domain, a Body Sensor Network (BSN) captures vital signs of patients for detecting emergencies. To determine the accurate health status of a patient, sensors are used. If the actual condition of the patient has deteriorated, new data and additional sensors need to be included.
This example illustrates a requirement change for the BSN, which has to be taken into account by runtime verification.
Moreover, requirements may generally change due to software evolution~\cite{Lehman1980}, which also applies to SAS~\cite{weyns2020introduction}. 

In this context, we advocate that \textit{runtime verification should be able to handle such requirement changes while checking the managed system against the requirements by preserving the knowledge it gathered until then and with minimal delay}. 
For instance, in the BSN, the runtime verification should preserve the knowledge about which of the existing sensors have already provided vital signs at the point in time when a sensor is deployed. Otherwise, the runtime verification would inaccurately require the existing sensors to send vital signs, which, however, they already did. Hence, a re-start of the runtime verification with an updated property could lead to inaccurate results and, thus, a period in which the patient is unassisted.  Thus, we aim for a co-evolution of the runtime verification and requirements, in which the verification seamlessly checks the managed system against  changing requirements. 

This proposal contrasts the state-of-the-art runtime verification approaches for SAS that do not support changing requirements in terms of dynamically adapting the properties~\cite{CailliauL19:TAAS,goldsby2007amoeba,filieri2011run,Calinescu2012quantitative, calinescu2017using, Solano+2019}. Other approaches provide only limited flexibility for requirements changes by relaxing requirements~\cite{Li:RE2022,Dippolito:ICSE14} or representing uncertainty in properties with fuzzy logic~\cite{whittle2010relax,baresi2010fuzzy}, but still without supporting adaptations of properties. 

In this paper, we propose a novel runtime verification technique for SAS that can seamlessly handle requirements changes by dynamically and safely adapting properties that formalize the requirements. Technically, each property is represented by an \textit{observer automaton} (or \textit{observer} for short) that has a dedicated error state denoting a violation of the property. To identify such a violation, the runtime verification traverses the observer in terms of state transitions triggered by monitored events emitted from the managed system or environment.
To cope with a changing requirement, an observer can be dynamically and safely adapted to reflect the changed requirement while the verification is running and the observer is quiescent (cf.~\cite{Kramer:TSE90}).
Thereby, intermediate verification results in terms of progress made in the observer is preserved to achieve a continuous, incremental verification as advocated by Ghezzi in the context of SAS~\cite{ghezzi2012evolution}.

To ease specification of properties, we leverage \textit{Property Specification Patterns} (PSP)~\cite{AutiliGLPT15} that support specifications in structured English (using a grammar) and automated translations to temporal logic. In our work, we rely on future MTL~\cite{MTL}.
Furthermore, we built upon our pattern catalog that provides templates for observers used for model checking real-time systems against Timed Computation Tree Logic (TCTL) properties at design-time~\cite{vogel2023property}. Based on these templates, we define novel templates for observers used for runtime verification concerning MTL properties. 
Thus, given a requirement in Structured English, we automatically generate the MTL property and the corresponding observer by using the translations of Autili et al.~\cite{AutiliGLPT15} and the novel observer templates. Therefore, the semantics of the observers are precisely defined to enable an accurate runtime verification.
To further provide precise semantics of observer adaptations, we propose in this paper \textit{Property Adaptation Patterns}~(PAP) that specify at the level of PSP how an observer should be adapted to accurately reflect changes of a requirement in the observer. For a concrete SAS, these PAP are instantiated to adaptation rules used by a higher-level feedback loop to dynamically adapt the observers used by the lower-level feedback loop for runtime verification. 

We evaluate our runtime verification technique on an Arduino prototype of the Body Sensor Network (BSN)~\cite{BSN} with changing requirements and the Timescales benchmark~\cite{ulus2019timescales}. Results show that our technique can cope with changing requirements while being efficient and correct.

Thus, this paper provides the following contributions:
\begin{enumerate}
    \item[C1)] An observer-based, incremental runtime verification technique that seamlessly copes with changing requirements.
    \item[C2)] A systematic approach using PSP to automatically obtain observers with precise semantics for runtime verification.
    \item[C3)] Property Adaptation Patterns (PAP) to precisely define observer adaptations and their semantics.
    \item[C4)] An implementation of our technique for the Arduino plattform and its evaluation on the BSN system and Timescales benchmark.
\end{enumerate}

The rest of the paper is structured as follows. In Section~\ref{sec:background}, we provide the background on the BSN, PSP, and observers. We discuss our runtime verification technique in Section~\ref{sec:approach}, evaluate it in Section~\ref{sec:evaluation}, and contrast it to related work in Section~\ref{sec:related}. Finally, we conclude our paper in Section~\ref{sec:conclusion}.

\section{Background}\label{sec:background}

\subsection{Running Example: The Body Sensor Network (BSN)}\label{sec:background:bsn}

Throughout this paper, we use the BSN~\cite{BSN} as our running example. The BSN can be used to monitor and analyze a patient's vital signs to determine if the patient is in an emergency state. If an emergency is detected, an emergency signal is sent to an external agent. 
To monitor the patient, the BSN can be equipped with sensors, such as a pulse sensor, a glucometer, or a thermometer. The \textit{BodyHub} acts as a central unit that requests data from the sensors to process it and evaluate the patient's risk status. 
The BSN is equipped with a scheduler that enables self-adaptive functionality to minimize energy usage while still providing sufficient confidence in the obtained data. 

In previous work~\cite{vogel2023property}, we expressed the requirements of the BSN with the help of PSP as TCTL properties and verified a model of the BSN with observer automata at design-time using the UPPAAL model checker~\cite{behrmann2004tutorial}. 

\subsection{Property Specification Patterns (PSP)}\label{sec:background:psp}

The formalization of requirements as properties expressed in a temporal logic requires expertise in the use of the logic. 
\textit{Property Specification Patterns}~(PSP) have been proposed for qualitative~\cite{DwyerAC99}, real-time~\cite{KonradCheng05, Gruhn06L}, and probabilistic~\cite{Grunske08} requirements, which all have been collected and extended in one unified catalog~\cite{AutiliGLPT15}. 
Such patterns define recurring schemes of requirements that are leveraged to ease the formalization.

Particularly, the catalog by Autili et al.~\cite{AutiliGLPT15} provides a grammar to express properties in Structured English language and mappings to various temporal logics such as LTL and MTL for an automated translation from Structured English to the logics. 
For instance, the timed \textit{Response} pattern is denoted in structured English as ``If $P$ has occurred, then in response $S$ eventually holds between $t_1$ and $t_2$'' while $P$ and $S$ are placeholders for events or states of a concrete system and $t_1$ and $t_2$ are variables defining a time window, in which $S$ should happen. The corresponding property is mapped to MTL as $\square (P \rightarrow \lozenge^{[t_1, t_2]} S)$ with the same placeholders.
Thus, using the grammar and translations to temporal logic, specification is eased and precise semantics of requirements is achieved.

\subsection{Automata-Based Runtime Verification}\label{sec:background:rv}

In general, automata-based approaches are one means to realize runtime verification since an automaton is an operational representation of a property defined declaratively in temporal logic~\cite{falcone2021taxonomy,bauer2011runtime}. 
We name such an automaton \textit{observer}. 
The states of an observer capture information about the past, and the transitions capture possibilities for the future behavior of the system under verification, which constitutes a \textit{past implies future} modality~\cite{falcone2021taxonomy}. Accepting or error states of an observer denote a satisfaction or violation of the property, and the active state represents the current state of the system.
Thus, a runtime verifier can traverse the observer based on system and environmental events to verify the behavior of the running system against the property.
Technically, the observer has to be created for a property expressed in temporal logic and can then be either directly interpreted by a verifier or further compiled down to code for execution within a verifier~\cite{falcone2021taxonomy}. 

In concrete approaches, observers are either generated manually~\cite{UMLObserver,Mallozzi+2019} or automatically from temporal logic~\cite{goldsby2007amoeba,Giannakopoulou+Havelund2001,bauer2011runtime} or other specifications such as sequence diagrams~\cite{Simmonds+2009}.

\section{Runtime Verification of Self-Adaptive Systems with Changing Requirements}\label{sec:approach}

In this section, we discuss our approach to runtime verification of SAS with changing requirements. 
First, we introduce the \textit{Property Specification and Adaptation Patterns} for runtime verification. Then, we detail how our approach uses these patterns in SAS for automata-based runtime verification, where requirements changes imply adaptations of properties and observers. We use the BSN as a running example.

\subsection{Property Specification and Adaptation Patterns}\label{sec:approach:catalog}

In our work, we focus on verifying SAS against properties expressed in MTL. To ease the formalization of requirements as MTL properties, we leverage the \textit{Property Specification Patterns} (PSP) from literature (Section~\ref{sec:background:psp}). We particularly reuse the Structured English Grammar and mapping to MTL formula templates from Autili et al.~\cite{AutiliGLPT15}. Thus, a user formalizes a requirement in structured English, which is automatically translated to an MTL property. 

Additionally, to realize an \textit{automata-based runtime verification}, we need to construct observers for MTL properties (Section~\ref{sec:background:rv}). To automate the construction of observers with PSP, we built upon our PSP catalog that provides observer templates (UPPAAL timed automata) for TCTL properties and focuses on design-time model checking with UPPAAL~\cite{vogel2023property}. However, direct reuse of these observer templates for runtime verification is not feasible, e.g., due to non-determinism in these observers and their focus on design-time model checking. 
Thus, we created new observer templates for MTL properties that are deterministic and focus on
runtime verification. Still, the existing templates have been a solid basis to obtain the new templates. These templates allow us to automatically construct observers for properties expressed in Structured English.

Our novel PSP catalog for runtime verification with its observer templates builds on the mapping from natural language to MTL~\cite{AutiliGLPT15} and observer techniques~\cite{vogel2023property}. Therefore, the catalog offers precise semantics in expressing properties and representing them in observers. 
In particular, we systematically created the observer templates by manually analyzing all possible types of traces that would violate a property, and generalizing these traces to a timed automaton. Thus, such an observer template represents a set of traces, of which some violate the corresponding property. To distinguish violating and satisfying traces, the observer template contains an \textit{error} state that is reached if and only if a trace violates the property.  

Additionally, with our catalog we propose \textit{Property Adaptation Patterns} (PAP) that define at the PSP level how observers should be adapted to represent changes of requirements. Thus, adaptations of properties are accurately reflected in the observers. Technically, PAP are defined by graph transformations (cf.~\cite{Giese+2012}) on observer templates that are instantiated to adaptation rules for concrete observers.
Thus, our catalog does not only provide precise semantics for specifying (using PSP) but also for adapting properties (using PAP).
The PSP/PAP catalog is publicly available\footnote{\url{https://github.com/hub-se/PAP/wiki}} and  detailed in the following sections.

\subsection{Architectural Overview}\label{sec:approach:architecture}

\begin{figure}
    \centering
    \includegraphics[width=.9\columnwidth]{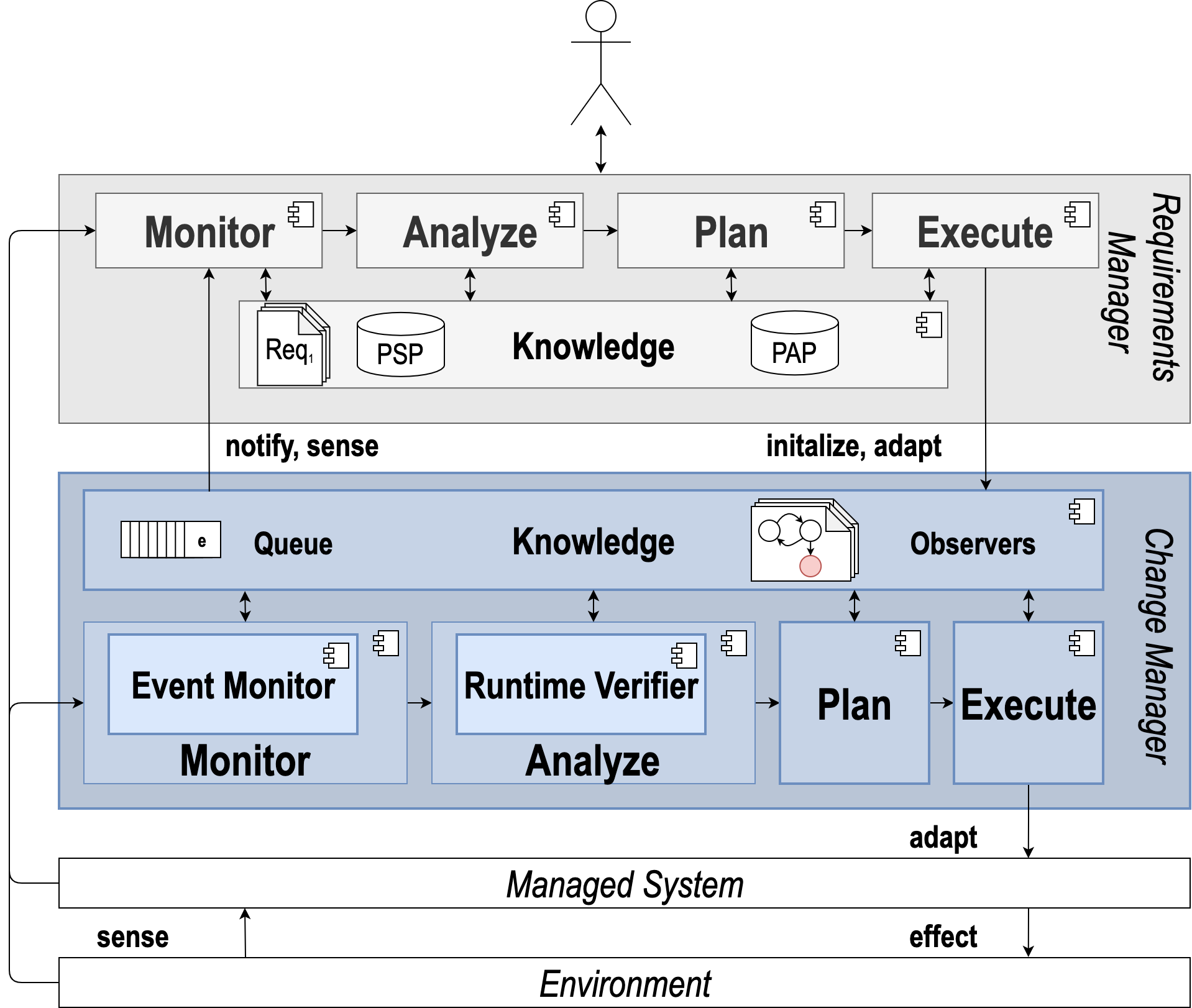}
    \caption{Architectural Overview.}
    \label{fig:architecture}
    \vspace{-2em}
\end{figure}

Fig.~\ref{fig:architecture} shows an architectural overview of our approach. We consider a SAS to be split into a managed and managing system operating in an environment. The managing system is split into two layers, each implementing a MAPE-K feedback loop: the \textit{Change Manager} and the \textit{Requirements Manager}.

\paragraph{Change Manager}
The change manager (shaded in blue in Fig.~\ref{fig:architecture}) adapts the managed system so that the system satisfies its requirements despite uncertainty. Adaptation is needed if requirements are violated. To determine such violations, the change manager performs \textit{automata-based runtime verification} (Section~\ref{sec:background:rv}).
For this purpose, it monitors the managed system and environment. The \textit{Event Monitor} adds events representing changes of the system and environment to the first-in-first-out \textit{Queue}. 
In the \textit{Analyze} step, the \textit{Runtime Verifier} consumes the events from the queue and matches them against the \textit{Observers}, each representing a property.

If a property is violated, adaptation of the managed system is needed so that the \textit{Plan} and \textit{Execute} steps are performed. 
The change manager interacts with the \textit{Requirements Manager} by notifications about property violations. The other way around, the requirements manager initializes the runtime verification by providing observers to the change manager and dynamically adapts these observers if requirements change.

\paragraph{Requirements Manager}
The requirements manager (top layer shaded in gray in Fig.~\ref{fig:architecture}) is in charge of formalizing requirements given by a human in Structured English to MTL properties and corresponding observers. It uses our \textit{Property Specification Pattern} (PSP) catalog comprising mappings to MTL and observer templates. The generated observers are provided to the change manager for runtime verification. 

At runtime, the requirements manager monitors the change manager, managed system, and environment to identify with the help of a human changes of requirements, for instance, triggered by a human or the change manager notifying about violations of requirements. 
In the \textit{Analyze} and \textit{Plan} steps, the requirements manager and human determine which requirements have changed and how they have changed to adapt the properties and corresponding observers used by the change manager accordingly. To accurately adapt properties and observers for runtime verification, our \textit{Property Adaptation Pattern} (PAP) catalog comprises adaptation templates for each observer template. These adaptation templates are instantiated to adaptation rules that are automatically and safely executed on the change manager's observers to reflect the changed requirement for runtime verification.
Changes of requirements may also mean that new requirements emerge or existing requirements become irrelevant. Thus, the requirements manager has to synthesize new properties and observers that are provided to the change manager or respectively remove existing observers from the change manager. 

We believe that fully automating the requirements manager, especially the analyze and plan steps, is challenging and also possibly not desired. Therefore, our current proposal includes the \textit{human in the loop} who is in charge of decision-making regarding identifying requirements changes and determining how properties/observers need to be adapted to reflect these changes. However, our PSP/PAP catalog supports the human by easing the specification and adaptation of properties with their observers while providing precise semantics for them. Moreover, the execution of dynamic and safe adaptations of properties/observers is performed automatically.

In the following, we detail how runtime verification is initialized and performed, and how properties are adapted.
In general, the system may have multiple requirements and each requirement may be expressed by multiple properties. For readability purposes, we describe in the following section how our approach handles one requirement expressed by one property. 
Nevertheless, our approach works with multiple requirements and multiple properties by deploying multiple independent instances of the \textit{Event Monitor}, including the \textit{Queue}, and the \textit{Runtime Verifier}.

\subsection{Initializing Runtime Verification}\label{subsec:initialzingRV}

\begin{figure}[t!]
    \centering
    \begin{subfigure}{0.5\textwidth}
        \centering
        \includegraphics[width=.9\columnwidth]{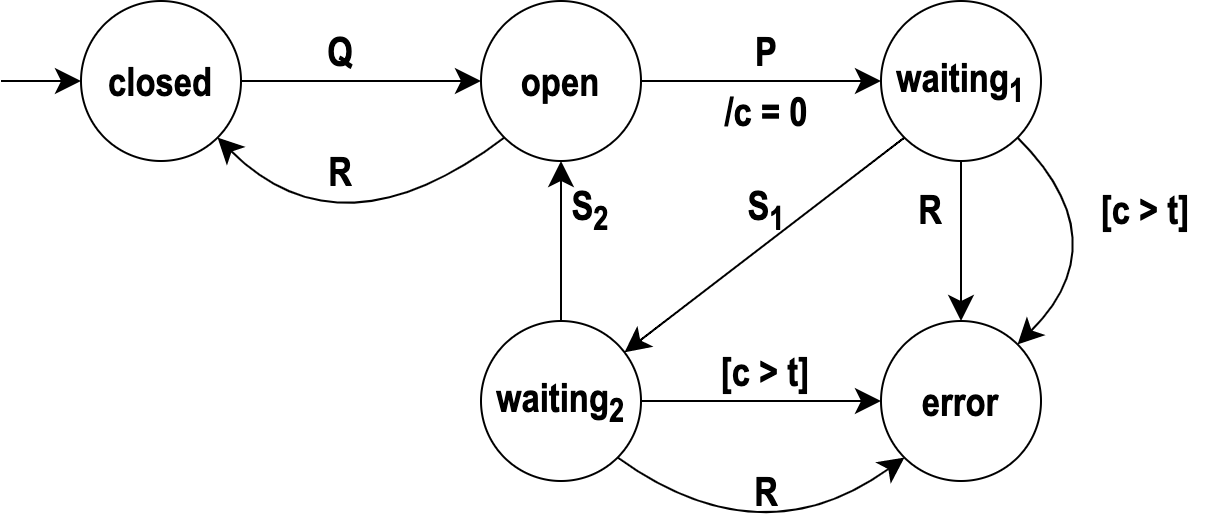}
        \caption{Observer template.}
        \label{fig:obs-template}
    \end{subfigure}%

    \begin{subfigure}{0.5\textwidth}
        \centering
        \includegraphics[width=.9\columnwidth]{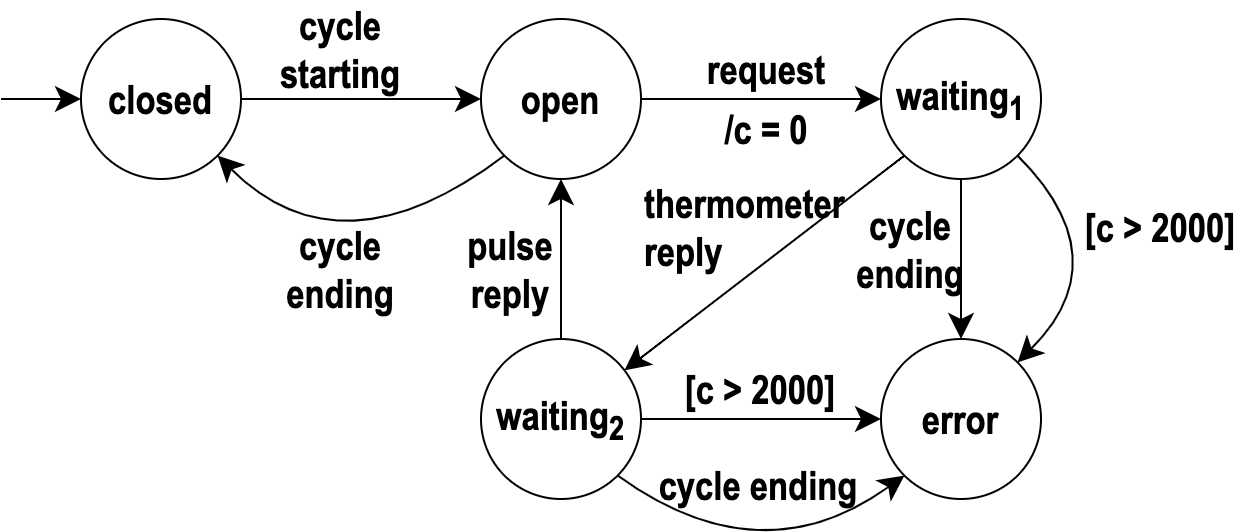}
        \caption{Instantiated observer from the template.}
        \label{fig:obs}
    \end{subfigure}%
    \caption{Observer for Timed Response Chain w/ Between scope. }
    \label{fig:obs-all}
    \vspace{-1em}
\end{figure}

Our PSP catalog in the \textit{Requirements Manager} is used when a stakeholder expresses a requirement in Structured English. The catalog then generates an MTL property formalizing the requirement. Moreover, the corresponding observer template is retrieved and instantiated to an observer that represents this property (an example is shown in Fig.~\ref{fig:obs-all}). The resulting observer is eventually provided to the \textit{Change Manager}'s \textit{Knowledge} and used by the \textit{Runtime Verifier} (cf. Fig.~\ref{fig:architecture}).

Conceptually, an observer contains a set of states, one of which is the current state representing the current state of the managed system and environment. Each of the observer's states has a set of outgoing transitions. Such a transition points to another state (the transition's \textit{target}) and may have a guard condition over clocks and an action to reset clocks. 
Additionally, transitions may be labeled with an event type.

After deployment, the runtime verifier sets the observer's current state to the initial state.
Additionally, based on the property, a list of event types relevant to evaluate the property is provided to the change manager's knowledge and used by the event monitor to filter relevant events emitted from the managed system and environment.\footnote{Technically, the managed system and environment do not need to emit events, but the monitor can sense the state of the system and environment and create corresponding events whenever relevant state changes occur.} Filtered events are added to the \textit{Queue}, from where they are processed for runtime verification by the runtime verifier (see Section~\ref{subsec:performingRV}).

\textbf{Example:}
In the BSN, the developer is aware that network congestions may arise. To this end, the change manager adapts the managed system to decrease network usage by reducing the number of scheduler cycles and therefore limiting how often the BodyHub requests data. Nevertheless, the developer wants to specify that in any case, both sensor nodes (thermometer and pulse sensor) shall send data to the BodyHub when the BodyHub requests such data. The developer sets a time limit of $2s$ within which both sensors shall respond. This behavior is expected to be repeated for every scheduler cycle. To formalize this requirement, the developer selects a suitable pattern (\textit{Timed Response Chain} with the \textit{Between} scope) from the PSP catalog and expresses the requirement using the structured English grammar:

\begin{quote}
    Between the scheduler cycle starting and elapsing, if \emph{the BodyHub requests data}, then in response \emph{the thermometer sends} eventually  \emph{within 2s }followed by \emph{the pulse sensor sends} \emph{within 2s}.
\end{quote}

The Timed Response Chain covers requirements that expect an ordered chain of events within a time window in response to a request event. The Between scope further requires that the request and responses together are surrounded by two events, in this case, describing the start and end of a scheduler cycle.
The requirements manager uses the PSP catalog to provide the MTL formula template
\begin{equation}
\begin{split}
        & \square (( Q \wedge \lozenge R ) \rightarrow \\ 
        & (P \rightarrow \linebreak (\neg R \mathcal{U}^{[0,t]} (S_1 \wedge \neg R \wedge (\lozenge ^{[0, t]} (S_2 ))  ))) \mathcal{U} R)
\end{split}
\end{equation}
as well as the corresponding observer template shown in Fig.~\ref{fig:obs-template} that both correspond to the selected pattern. Both templates are instantiated and the placeholders $P, Q, R, S_1, S_2$ are replaced with actual values from the Structured English requirement. This results in the following MTL formula:
\begin{equation}\label{eq:req1}
\begin{split}\raisetag{2.8em}
        & \square (( \text{cycle\_starting} \wedge \lozenge \text{cycle\_ending} ) \rightarrow (\text{request} \rightarrow \\
        & (\neg \text{cycle\_ending} \text{ }\mathcal{U}^{[0,2]} (\text{thermometer\_reply} \text{ }\wedge \\ 
        & \neg \text{cycle\_ending} \wedge (\lozenge ^{[0, 2]} (\text{pulse\_reply} ))  ))) \text{ }\mathcal{U}\text{ } \text{cycle\_ending})
\end{split}
\end{equation}
and the observer shown in Fig.~\ref{fig:obs}.
 
In the observer, there are states corresponding to the scheduler cycle not having started (\emph{closed}), the cycle having started (\emph{open}), the BodyHub requesting data (\emph{waiting$_1$}), and the first sensor, but not yet the second sensor sending data (\emph{waiting$_2$}). Additionally, there is an \emph{error}-state that is reached if and only if the scheduler cycle elapses before both sensors send data, or if the time bound of $2s$ elapses. 

The event types that need to be monitored to evaluate the property are: (i) the scheduler cycle starts, (ii) the scheduler cycle ends, (iii) the BodyHub requests data, (iv) the pulse sensor sends, and (v) the thermometer sends.
The observer's current state is set to its initial state, which is \emph{closed}.

\subsection{Performing Runtime Verification}\label{subsec:performingRV}

Once the observer is deployed to the change manager, the \textit{runtime verifier} traverses the observer to verify the managed system against the property encoded in the observer. To this end, it checks if outgoing transitions of the observer's current state are enabled. 
A transition is enabled if its guard condition is evaluated to \textit{true}. In general, our observers only have guard conditions that refer to clock valuations against time bounds. Therefore, an observer has a clock that can be reset when a transition is taken.  
Transitions equipped with a label are only enabled when an event instance of the labeled event type is processed. 
Disabled transitions with a guard condition that are unlabeled may become enabled simply because time progresses. To take such a transition as soon as the guard's valuation changes from \textit{false} to \textit{true}, the runtime verifier analyzes all such transitions starting in the observer's current state upon entering it and determines the amount of time that needs to pass for each transition to become enabled. For the smallest such time, a timer is set that will generate an event that triggers the runtime verifier to force progress in the observer. Therefore, the runtime verifier only needs to access the observer when an event is monitored, either stemming from the managed system or environment, or from an elapsed timer. If the state in the observer is switched due to events from the managed system or environment and before the timer elapses, the timer is discarded.

To perform runtime verification, the \textit{event monitor} observes the managed system and environment and puts the events determined as relevant in the first-in-first-out event queue, which serves as a buffer of events, maintained by the knowledge component.\footnote{For runtime verification, we assume that monitored events from the managed system and environment have a strict order, that is, one event is monitored after another. Thus, an event is a tuple of an instance of some general event type and a timestamp. A trace is a list of such events.} The \textit{runtime verifier} processes these events from the queue one after the other. For each consumed event, it checks whether any outgoing transition of the observer's current state is enabled. 
If a transition is enabled, it is taken and the observer's current state is set to the transition's target. 
Otherwise, when no transition is enabled, the observer remains in its current state. At this point, the processed event is discarded. 
Upon reaching a new state, again all outgoing transitions are checked. If no transition is enabled, the observer remains in its current state and a timer is set according to the outgoing transitions' guard conditions (if applicable). 
Therefore, while the observer's structure represents the property, its current state represents previously obtained knowledge about the execution of the managed system and its environment until now.

If and only if the observer reaches an error state during the runtime verification, the managed system in its environment violates the property. 
In the change manager, detecting such a violation can act as a stimulus that, among others, triggers it to plan and execute an adaptation of the managed system. The change manager may also notify the requirements manager of the violation.
We designed this observer-based verification technique to be used in an online setting, that is, the verification is performed alongside the running managed system and environment that provide a continuing stream of events. However, the technique can be used without modifications for offline verification when a trace of events is provided later. 

\textbf{Example:}
Suppose the following execution of the BSN regarding the property specified above: 
First, a scheduler cycle starts. The monitor adds the corresponding event to the queue before it is analyzed. For the observer's current state, there is an enabled transition for the monitored event. The transition is taken and the observer progresses to state \emph{open}. $100ms$ later, the \emph{BodyHub requests} data from the sensors. Processing this event, the observer is progressed to state \emph{waiting$_1$} and its clock $c$ is reset. Upon analyzing the current state's outgoing transitions, a timer is set to $2,000ms$. If the thermometer and pulse sensor do not send data in return before the timer elapses, the observer progresses to the \textit{error} state. 
During the BSN's execution, the change manager may perform adaptations in the scheduler, that is, the number of scheduler cycles may be decreased to improve confidence in the obtained data, or increased to reduce energy consumption and network usage.

\subsection{Adapting the Property During Runtime Verification}\label{sec:approach:pap}
At runtime, the requirements manager monitors the change manager, managed system, and environment to identify and handle requirements changes with the help of a human. 
It can further react to notifications from the change manager that the managed system currently violates the property. 
If the requirements manager finds that the previously specified property is no longer adequate, it utilizes the \textit{property adaptation patterns}~(PAP) to systematically adapt the existing property. These PAP extend our PSP catalog to define adaptations of properties and observers at the pattern level (Section~\ref{sec:approach:catalog}). Therefore, they provide precise semantics for such adaptations.

Particularly, the requirements manager selects the PAP that appropriately reflects the requirements change in the property and observer, instantiates the PAP to an adaptation rule, and applies this rule to dynamically adapt the observer deployed in the change manager. 
Thus, the observer representing the requirement seamlessly co-evolves with the requirement, which contrasts discarding and redeploying a new observer in the case of requirements changes.
Therefore, the observer's current state can persist through adaptation, which is beneficial as it reflects information obtained previously about the execution of the managed system and environment until the adaptation. This leverages an incremental verification (cf.~\cite{ghezzi2012evolution}) where previous knowledge is preserved for the runtime verification. 

Nevertheless, enacting an adaptation of the observer has to be synchronized with the runtime verification that uses the same observer so that the adaptation is \textit{safe}. Otherwise, adapting the observer while the runtime verifier traverses the observer and performs state transitions could lead to inconsistencies. To achieve safe adaptations, the observer can only be adapted when it is quiescent (cf.~\cite{Kramer:TSE90}). 
Therefore, the requirements manager adds a dedicated \textit{adaptation event} to the event queue of the change manager. When this event is processed by the runtime verifier, the adaptation of the observer is performed instead of a verification step. After the adaptation, the runtime verifier continues processing the monitored events and performing verification steps. This approach also ensures that all events monitored before the adaptation event are processed with the unchanged observer.
 
Since our observers are based on PSP, both the original and adapted property can be expressed in structured natural language to describe the adaptation. The PAP range from \textit{parameter} (i.e., updating time bounds or replacing event types corresponding to placeholders in MTL formula templates of PSP) to \textit{structural} adaptations (i.e., the structure of the underlying property and observer are adapted, e.g., by adding or removing a response in a response-chain property resulting  in novel or obsolete states in the observer).

In the following, we present five PAP. We outline them in natural language and formalize exemplarily two of them with graph transformations on observers.
For these PAP, we noticed that a seamless adaptation of the observer preserving the already obtained knowledge in contrast to redeployments and restarts of verification processes is beneficial. 
Still, we do not claim that there is no situation in which a redeployment and restart of the observer can be appropriate.

We present the following five PAP: 
a) updating a time guard,
b) updating an event,
c) adding a response to a chain,
d) removing a response from the chain,
and e) splitting the response chain into multiple response properties.
While patterns (a) and (b) cover parametric changes of the requirement, patterns (c), (d), and (e) cover structural changes.

\paragraph{Updating a Time Guard}
This PAP covers changes of a deadline in a real-time requirement by adapting a property's time guard. 
Such an adaptation is performed by updating the corresponding guards in the observer and all timers. Adapting a time interval might change the valuation of guard conditions, and therefore enable previously disabled transitions. Thus, adapting a time interval might yield an immediate violation of the property. 
This PAP can be applied in this fashion to real-time properties of most patterns from the PSP catalog such as the Response, Existence, Absence, and Recurrence. 

\paragraph{Updating an Event}
This PAP is used to exchange one of the events specified in the property. In the observer template, such an adaptation can be performed by updating the labels of transitions from their old value, such as \textit{P}, to their new values, such as \textit{P'}.
This PAP can be applied to any pattern and observer in our catalog.

\paragraph{Adding a Response to the Chain}

\begin{figure}[t!]
    \centering
    \begin{subfigure}{0.5\textwidth}
        \centering
        \includegraphics[width=.7\columnwidth]{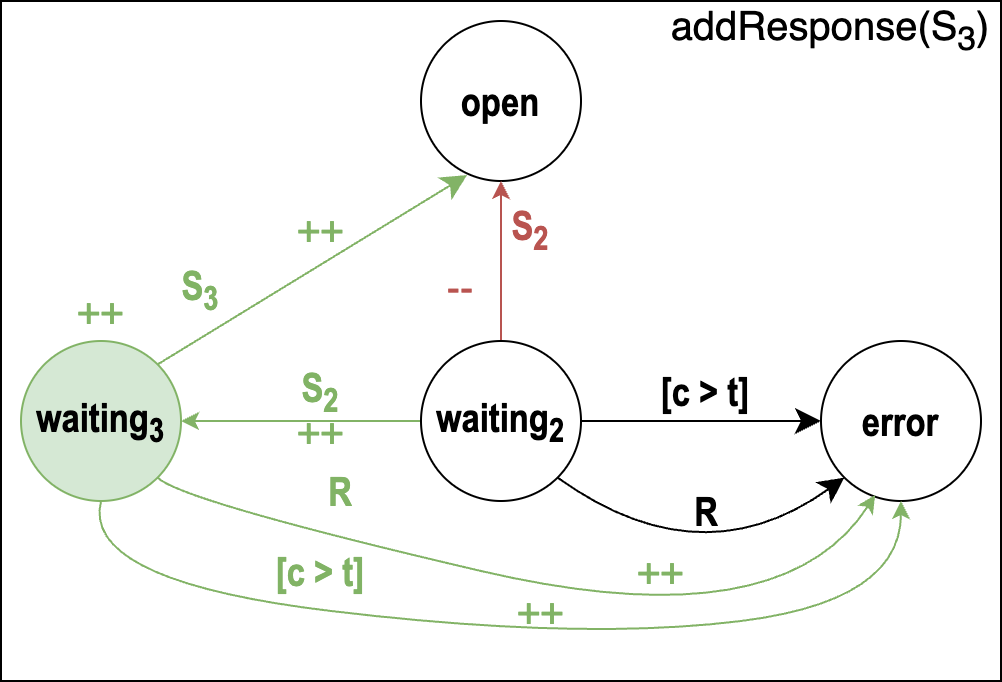}
        \caption{Graph transformation rule for the PAP.}
        \label{fig:obs-template-add}
    \end{subfigure}%

    \begin{subfigure}{0.5\textwidth}
        \centering
        \includegraphics[width=.9\columnwidth]{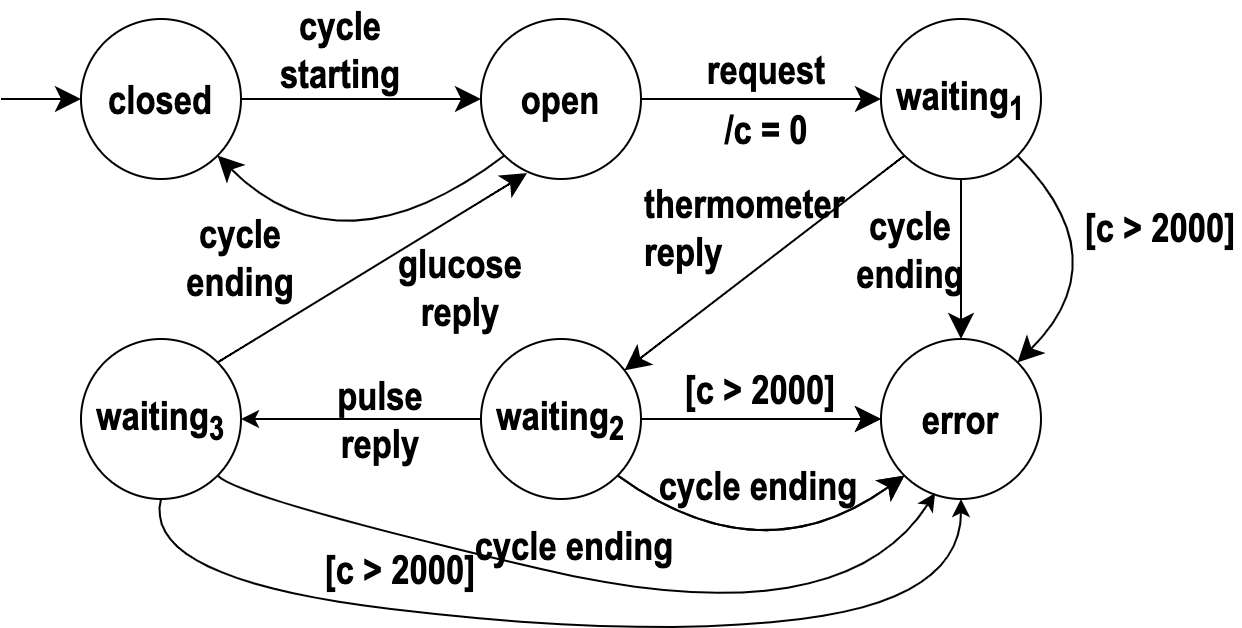}
        \caption{Resulting observer after adding a glucose sensor.}
        \label{fig:obs3}
    \end{subfigure}%
    \caption{PAP of adding a response to a response chain.}
    \label{fig:obs-templates}
    \vspace{-2em}
\end{figure}

This PAP is used when an additional response is expected to occur in the chain of responses, which extends the property. 
In essence, the response chain pattern defines a list of responses. The PAP allows the addition of a response to the end of the list, which is defined by the graph transformation rule (cf.~\cite{Giese+2012}) shown in Fig.~\ref{fig:obs-template-add}. The response to be added is defined by the rule's parameter $S_3$. This PAP can be applied to the observer template of the Response Chain pattern shown in Fig.~\ref{fig:obs-template} by matching the black and red elements of the rule in the observer template and afterwards performing the side effects of the rule, that is, removing the red elements (that are further annotated with $--$) and adding the green elements (that are further annotated with $++$) to the observer template. Accordingly, a new state $waiting_3$ is added to the observer template during which the new response $S_3$ is expected after response $S_2$ has occurred to move to state \textit{open}, otherwise to move to the \textit{error} state if the time bound has passed. 

However, the graph transformation rule shown in Fig.~\ref{fig:obs-template-add} is mainly considered as a specification of an observer adaptation at the pattern/template level that guarantees precise semantics of the adaptation according to the PSP. 
Thus, in a SAS, this rule is not applied on an observer template but it is rather itself a template. It will be instantiated for adapting a concrete property and observer. Instantiating and applying the rule to the observer of the BSN shown in Fig.~\ref{fig:obs} results in the observer shown in Fig.~\ref{fig:obs3}. Particularly, this adaptation reflects the requirements change that the BSN has to consider the data sensed and sent as \textit{glucose reply} by the glucose sensor, which have not been considered before. The response \textit{glucose reply} is expected to occur as the last response of the chain.

The observer's current state persists through the adaptation process, that is, the runtime verifier preserves the knowledge of which other responses of the chain have already occurred.

\begin{figure}[t!]
    \centering
    \begin{subfigure}{0.45\textwidth}
        \centering
        \includegraphics[width=.7\columnwidth]{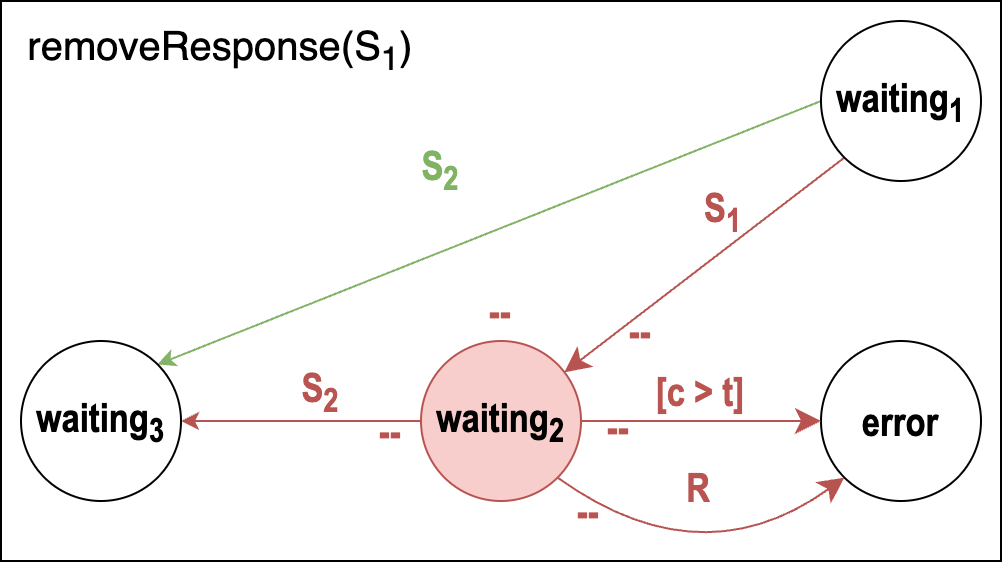}
        \caption{Graph transformation rule for the PAP.}
        \label{fig:obs-template-rem}
    \end{subfigure}%
     
    \begin{subfigure}{0.45\textwidth}
        \centering
        \includegraphics[width=.9\columnwidth]{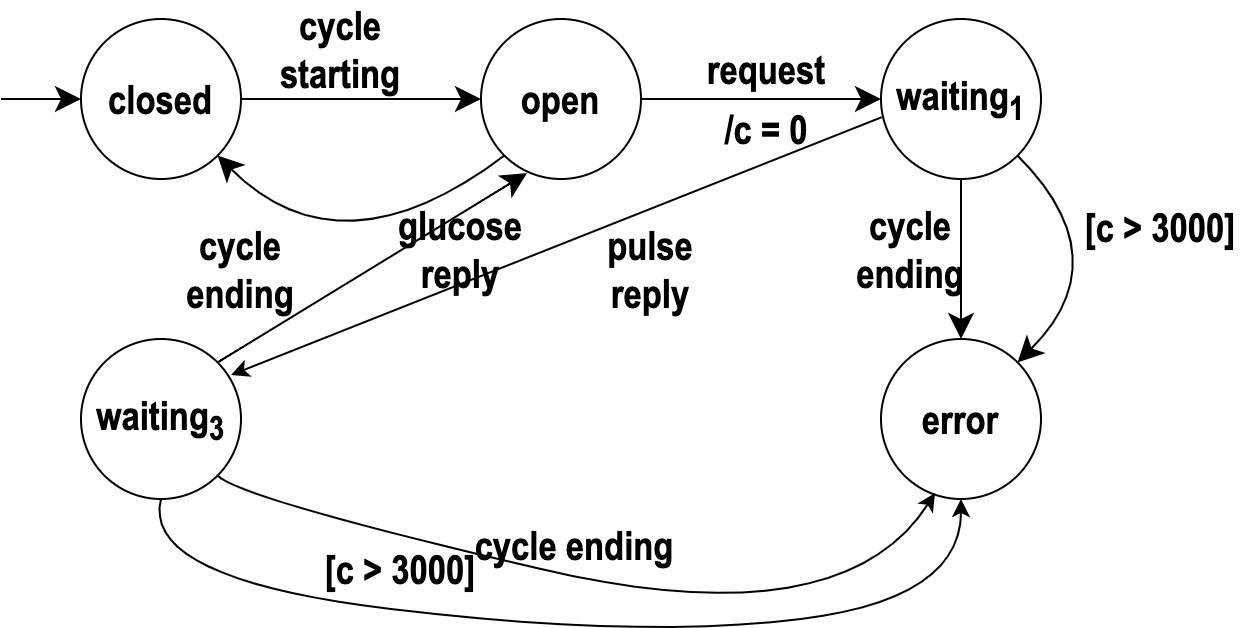}
        \caption{Resulting observer after removing a thermometer.}
        \label{fig:obs2'}
    \end{subfigure}%
    \caption{PAP of removing a response from the chain.}
    \vspace{-2em}
\end{figure}

\paragraph{Removing a Response from the Chain}
This PAP is used when an existing response is not needed anymore in the chain.
It is the counterpart of the previous PAP (adding a response).  
We now consider the removal of a response $S_1$ in the middle of the chain as shown by the graph transformation rule in Fig.~\ref{fig:obs-template-rem}. 
The rule is instantiated and applied similarly to the rule shown in Fig.~\ref{fig:obs-template-add}, but in this case to adapt the observer shown in Fig.~\ref{fig:obs3} to obtain the observer shown in Fig.~\ref{fig:obs2'}. This adaptation reflects the requirements change in the BSN that the thermometer is not needed anymore and therefore, the \textit{thermometer reply} is not expected to happen anymore.  

However, since this adaptation removes a state from the observer, we have to take into account that the state to be removed can be the \textit{current} state of the observer. If this is the case, a new current state has to be determined and set by the adaptation. 
If the current state is $waiting_i$ that should be removed by adaptation, we know that the property's scope is open (scheduler cycle has started), a request has occurred, and all responses prior to $S_i$ have happened.
In the adapted observer, the same information is represented by state $waiting_{i+1}$, since the state $waiting_i$ together with the response $S_i$ have been removed by adaptation. Thus, in the case that the \textit{current} state was removed, the current state of the observer after adaptation is set to $waiting_{i+1}$.

\paragraph{Splitting the Response Chain}
In a response chain, the order of expected responses is specified. If a specific order of the responses is no longer required, this PAP splits the chain into multiple, independent responses, that is, multiple response chains with only one response for each chain. Therefore, it generates multiple observers, one for each response, and maintains the information represented by the existing observer by selecting the current state for each of those new observers depending on the current state of the existing observer. 

Consider, for example, a response chain with two responses as shown in Fig.~\ref{fig:obs}. The observer's state \emph{closed} represents that the scope is closed, \emph{open} represents that the scope is open but no response is required because any previous request has already been replied, \emph{error} represents a violation of the property. Each of these three states can also be found in the new observers, where they still represent the same information. Hence, if the existing observer is in any of these states, the new observers will be set to that state as well. 
If the existing observer's current state is $waiting_i$, i.e., any of the states where a request has been sent but not all replies have occurred, the information for the new observers differs according to the response they are representing. Consider, for example, that $waiting_2$ is the current state in the existing observer. This means that the first response has occurred, but not yet the second. Thus, the new observer for the first response should have \emph{open} as the current state, representing that all requests have been addressed by a response, while the other new observer (for the second response) should have \emph{waiting} as its current state because a request has occurred but not yet the corresponding response (i.e., $S_2$). In general, all observers representing properties regarding responses $S_j$ with $j < i$ will have their current state set to \textit{open}, while the remaining observers will be in state \textit{waiting}. 

\smallskip

After an observer adaptation regardless of the PAP that is used, the runtime verifier checks all outgoing transitions of the current state in the observer since any adaptation may enable previously disabled transitions. Additionally, existing timers are discarded and new timers are set accordingly for updated timed guards in the observer. Afterwards the runtime verifier continues with regular verification steps by consuming monitored events from the queue (see Section~\ref{subsec:performingRV}). 

\section{Evaluation}\label{sec:evaluation}

We evaluate our observer-based runtime verification approach by investigating the following research questions: 
\begin{itemize}
    \item [RQ1:] How efficient is the observer-based runtime verification in terms of time for processing monitored events with observers and memory needed to represent observers?
    \item [RQ2:] How accurate is the observer-based runtime verification in detecting violations of properties? 
    \item [RQ3:] How fast is a dynamic adaptation of an observer at runtime compared to a redeployment of the observer? 
    \item [RQ4:] Can the observer-based runtime verification with adapting observers increase the trustworthiness of SAS?
\end{itemize}

To perform our evaluation, we implemented our runtime verification approach with its  observers and adaptations following the PSP and PAP with C++ on Arduino.\footnote{The PSP/PAP catalog, implementation, and replication package for the evaluation are available at: \url{https://www.github.com/HUB-SE/PAP/}} 
We deployed the code on multiple Arduino Mega\footnote{Arduino Mega 2560 Rev3, 8 KB SRAM, 16MHz clock speed.}. 

\noindent
\textbf{RQ1 } 
This research question addresses the efficiency of our observer-based runtime verification. To perform verification online, the runtime verifier is desired to process monitored events faster than the managed system and environment emit them. Thus, the verifier can provide fast results without the possibility of an overflowing event queue. 

To determine how much time it takes for our observer-based verification technique to process events, we generated ten artificial traces containing events of five different types. The traces have a length of $50,000$ events. We measured the time that our technique took with an artificial observer to process these traces. This artificial observer contains five states. Each state has five outgoing transitions, each labeled with one of the five event types.\footnote{We omit evaluating the costs of managing timers due to guarded transitions in the observer because they are similar to processing an event requiring in both cases to check all outgoing transitions of the observer's current state.} Thus, with each processed event of the trace, one transition will be enabled regardless of the current state of the observer.
For each processed event, any outgoing transition of the current state has to be checked until the transition with the matching label is found. Overall, the artificial observer has 25 transitions, which  in our experience is a realistic upper bound for an observer~\cite{vogel2023property}. 

We execute our runtime verification technique with the artificial observer on Arduino Mega against the ten artificial traces. On average, our technique took $6571.1ms$ to process a trace with a standard deviation of $\mu$\,$\leq$\,$7.2ms$.
Thus, on average it takes $0.13ms$ ($6571.1ms$/$50,000$ events) for our technique to process a single event with an observer. 
Prominent benchmarks for runtime verification such as Timescales~\cite{ulus2019timescales} provide traces that contain one event per \textit{ms}. Thus, we conclude that our observer-based runtime verification technique is sufficiently efficient concerning the execution time. This especially holds since we check only monitored events representing changes of the managed system or environment, where we consider a rate of one change per $ms$ as extraordinarily high with respect to our experience with the BSN. 

We also investigate the memory needed to represent an observer in a data structure implemented on Arduino in terms of SRAM usage. For this purpose, we use observers for nine properties with different combinations of patterns and scopes as well as the artificial observer discussed previously.
As shown in Table~\ref{tab:memory}, the observers consume between \textit{355} and \textit{1,136 bytes} of memory. For each observer, we list the number of states and transitions to illustrate the size of the observer. Such sizes are representative of properties following the PSP. 
We conclude that observers can be efficiently represented given their size in terms of states and transitions and multiple observers each representing a property can be deployed to one Arduino Mega that has \textit{8KB} of SRAM. 
\begin{boxA}
    Our runtime verification is efficient as it just requires $0.13ms$ on average to process a monitored event and between $355$ and $1,136$ \textit{bytes} of memory to represent an observer. 
\end{boxA}

\begin{table}[tbp]
    \begin{center}
        \caption{Size of observers for properties in terms of numbers of states (\#S) and transitions (\#T), and memory usage in bytes.}
        \label{tab:memory}
        \vspace{-.5em}
        \begin{tabular}{c c c c}
            \toprule
            \textbf{Pattern + Scope} & \textbf{\#S} & \textbf{\#T} & \textbf{Memory (bytes)}\\
            \midrule
            Absence After & 5 & 4 & 614 \\
            Absence Before & 5 & 4 & 558 \\
            Absence Between & 6 & 8 & 866 \\
            Recurrence Globally & 2 & 2 & 355 \\
            Recurrence Between & 4 & 5 & 605 \\
            Response Globally & 3 & 3 & 458 \\
            Response Between & 4 & 6 & 652 \\
            Response Chain Between, 2 responses & 6 & 11 & 940 \\
            Response Chain Between, 3 responses & 7 & 14 & 1,136 \\ \midrule
            Artificial observer & 5 & 25 & 1,047 \\
            \bottomrule
            \end{tabular}
    \end{center}
    \vspace{-2.25em}
\end{table}

\noindent
\textbf{RQ2 } 
In this research question, we investigate the correctness of our runtime verification. To this end, we implemented a trace generator according to the grammar of Timescales~\cite{ulus2019timescales}, which is a runtime verification benchmark. Each trace targets a property based on the PSP catalog and can be generated to either satisfy or violate the property. Thus, the generator provides the ground truth of whether a generated trace violates or satisfies the property.
We considered nine properties that follow the patterns and scopes shown in Table~\ref{tab:memory}.
For each property, we generated 20 different traces, each consisting of about 60 events. Ten of them satisfy and ten violate the property. 
Afterward, we instantiated the observer template of our PSP catalog for the property. We deployed the resulting observer and evaluated, whether it reaches an \textit{error} state when processing the trace. 
We found that our observers classified each of the 20 traces correctly for each of the nine properties. 
\begin{boxA}
    We can report 100\% accuracy in detecting property violations since our observer-based runtime verification has provided correct results for all 180 runs of the experiment (20 traces for each of the nine properties).
\end{boxA}

\begin{table*}[tbp]
    \begin{center}
        \caption{Requirements changes for the BSN. Added/updated parts of the property along the changes are highlighted in blue.}
        \label{tab:scenarios}
        \begin{tabular*}{\textwidth}{c L{0.12\textwidth} L{0.13\textwidth} L{0.67\textwidth}}
            \toprule
            \textbf{\#} & \textbf{Req. Change} & \textbf{PAP} & \textbf{MTL Property} \\
            \midrule
            0 & Initial situation (cf. Eq.~\ref{eq:req1}) & -- &$\square (( \text{cycle\_starting} \wedge \lozenge \text{cycle\_ending} ) \rightarrow (\text{request} \rightarrow (\neg \text{cycle\_ending } \text{ }\mathcal{U}^{[0,2]}\text{ } (\text{thermometer\_reply} \wedge \linebreak \neg \text{cycle\_ending} \wedge (\lozenge ^{[0, 2]} (\text{pulse\_reply} ))  ))) \text{ }\mathcal{U}\text{ } \text{cycle\_ending})$\\ \midrule
            1 & Add a glucometer & Adding a Response to the Chain & $\square (( \text{cycle\_starting} \wedge \lozenge \text{cycle\_ending} ) \rightarrow (\text{request} \rightarrow  (\neg \text{cycle\_ending} \text{ }\mathcal{U}^{[0,2]}\text{ }  (\text{thermometer\_reply} \wedge \linebreak \neg \text{cycle\_ending} \wedge (\lozenge ^{[0, 2]} (\text{pulse\_reply} )) \textcolor{blue}{\wedge \neg \text{cycle\_ending} \wedge (\lozenge ^{[0, 2]} (\text{glucose\_reply} ))} ))) \text{ }\mathcal{U} \text{ }\text{cycle\_ending})$\\ \midrule
            2 & Update time guard & Updating a Time Guard & $\square (( \text{cycle\_starting} \wedge \lozenge \text{cycle\_ending} ) \rightarrow (\text{request} \rightarrow (\neg \text{cycle\_ending} \text{ }\mathcal{U}^{[0,\textcolor{blue}{3}]}\text{ } (\text{thermometer\_reply} \wedge \neg \text{cycle\_ending} \wedge  (\lozenge ^{[0, \textcolor{blue}{3}]} (\text{pulse\_reply} )) \wedge \neg \text{cycle\_ending} \wedge (\lozenge ^{[0, \textcolor{blue}{3}]} (\text{glucose\_reply} )) ))) \text{ }\mathcal{U}\text{ } \text{cycle\_ending})$\\ \midrule
            3 & Remove the thermometer & Rem. a Response from the Chain & $\square (( \text{cycle\_starting} \wedge \lozenge \text{cycle\_ending} ) \rightarrow (\text{request} \rightarrow  \neg \text{cycle\_ending} \wedge  (\lozenge ^{[0, 3]} (\text{pulse\_reply} )) \wedge \neg \text{cycle\_ending} \wedge (\lozenge ^{[0, 3]} (\text{glucose\_reply} )) ) \text{ }\mathcal{U}\text{ } \text{cycle\_ending})$\\ \midrule
            4 & Scheduler requests data & Updating an Event & $\square (( \text{cycle\_starting} \wedge \lozenge \text{cycle\_ending} ) \rightarrow (\text{\textcolor{blue}{s\_request}} \rightarrow  (\neg \text{cycle\_ending} \text{ }\mathcal{U}^{[0,3]}\text{ } (\text{pulse\_reply} \wedge \neg \text{cycle\_ending} \wedge (\lozenge ^{[0, 3]} (\text{glucose\_reply} ))  ))) \text{ }\mathcal{U} \text{ }\text{cycle\_ending})$\\ \midrule     
            5 & Neglect order of sensors & Splitting the Response Chain & $\square (( \text{cycle\_starting} \wedge \lozenge \text{cycle\_ending} ) \rightarrow (\text{s\_request} \rightarrow  (\neg \text{cycle\_ending} \text{ }\mathcal{U}^{[0,3]}\text{ } (\text{pulse\_reply}))) \text{ }\mathcal{U}\text{ } \text{cycle\_ending})$ -- \textit{and a similar property for {glucose\_reply}}\\

            \bottomrule
            \end{tabular*}
    \end{center}
    \vspace{-2em}

\end{table*}

\noindent
\textbf{RQ3 }
This research question addresses the performance of a dynamic adaptation of a property at runtime. Thus, we compare the runtime efficiency of a \textit{dynamic adaptation} based on our PAP and a \textit{redeployment} of an observer. A redeployment comprises invoking the destructor to free up the memory consumed by the observer and the constructor to instantiate and represent the new observer in freshly allocated memory. 
For this experiment, we use the Response Chain property shown in Eq.~\ref{eq:req1}. For requirements changes, we alternate between adding and removing responses from the chain as well as updating response events in the chain. Such changes can easily be repeated multiple times on an observer to achieve reliable time measurements. For one run, we alternate between the changes until each of them is performed $1,000$ times resulting in a total of $3,000$ changes that are either realized by $3,000$ dynamic adaptations or $3,000$ redeployment of the observer. We repeat both runs ten times. The runs are all executed on Arduino.

On average across the ten runs, the $3,000$ dynamic adaptations of the observer took in total $3.034s$ (stdev $\mu_1$\,$\leq$\,$0.49ms$). 
Thus, one dynamic adaptation of an observer takes on average \textit{1.01ms}. 
In contrast, the $3,000$ redeployments of the observer took on average $15.330s$ (stdev $\mu_2$\,$\leq$\,$0.85ms$), that is, on average \textit{5.11ms} for one redeployment of an observer. 

\begin{boxA}
    A dynamic adaptation of an observer ($1.01ms$) is more than five times faster than a redeployment ($5.11ms$). 
\end{boxA}

\noindent
\textbf{RQ4 }
For the last research question, we investigate how our approach of dynamically adapting observers to reflect requirements changes can increase the trustworthiness of a SAS. To this end, we use a port of the BSN artifact~\cite{BSN} we implemented for the Arduino platform.
Starting with an initial situation of the BSN described by the requirement discussed in Section~\ref{subsec:initialzingRV} and formalized by the MTL property in Eq.~\ref{eq:req1}, we consider a sequence of five requirements changes shown in Table~\ref{tab:scenarios}. For each requirements change, the table shows the PAP to specify the adaptation of the property/observer and the MTL property after adaptation. We execute the BSN alongside our runtime verification approach and use the PAP to specify and perform the dynamic adaptations of the observer to reflect sequentially these five requirements changes in the verification.

With this demonstration of our approach, we show that using PAP allows us to specify adaptations of observers with precise semantics as shown by the corresponding MTL properties before and after an adaptation (cf.~Table~\ref{tab:scenarios}). Such adaptations of observers are dynamically and safely performed so that our approach preserves the knowledge---in terms of intermediate verification results as progress made in the observer---without compromising the integrity of the observer. 
For the given property that is adapted (Table~\ref{tab:scenarios}), the knowledge preserved in the observer comprises whether a scheduler cycle has already started and if so, which of the sensors already have and which still need to send data to the BodyHub. 
Thus, our approach achieves an incremental, continuous verification of the currently executing scheduler cycle against the adapted property.
Without preserving this knowledge (e.g., by a redeployment), the currently executing scheduler cycle remains unverified against the adapted property as the observer is reset to its initial state where it expects a novel cycle to start (cf.~\textit{cycle\_starting} event). In such a situation, there is no verification evidence about the safety of the BSN. 
\begin{boxA}
    Applying the PAP enables a continuous, incremental verification of the BSN that increases the trustworthiness of the BSN when requirements changes occur. 
\end{boxA}

\noindent
\textbf{Discussion}
In our evaluation, we have shown the efficiency of our observer-based runtime verification ($0.13ms$ to process a monitored event and at most $1.136$ bytes to represent a monitor) and adaptation of properties based on PAP ($1.01ms$ to dynamically adapt an observer). Thus, our approach can efficiently be used on microcontrollers such as Arduino. 

Moreover, we have shown empirically the correctness of our observer-based runtime verification using a benchmark based on Timescales~\cite{ulus2019timescales} as ground truth. Since there is no ground truth for verifying a running system against adapted properties, we cannot validate the correctness of our verification approach under changing requirements. Thus, we demonstrated qualitatively the benefits of continuous, incremental runtime verification on the trustworthiness of the~BSN~\cite{BSN}. 

\noindent
\textbf{Threats to Validity}
Threats to the validity of our study are as follows. 
\textit{Construct:}
Potential errors in our implementation of the observers and the Timescales grammar cause a threat to the validity of our reported results on correctness. We address this threat by having reviewed the observers and making the implementation and replication package publicly available. 
\textit{Internal:}
Threats of this category concern the experiments and measurements we conducted. To mitigate measurement errors and obtain reliable results, we repeated experiments and performed them on a SEAMS artifact ported to Arduino and on a benchmark based on Timescales that is used by runtime verification research community. 
Moreover, requirements formalized with the Structured English Grammar~\cite{AutiliGLPT15} might not match the stakeholders' intentions, which is also true for the properties/observers and eventually for the verification results. In this context, we rely on our expertise on the BSN~\cite{BSN,Solano+2019} and property specification patterns~\cite{AutiliGLPT15,vogel2023property}.  
\textit{External:}
We considered only the BSN with one requirement that changes in five ways in our study. Thus, our results may not generalize to other SAS, requirements, and changes.
Finally, our PSP/PAP catalog currently supports four PSP with different scopes and five PAP, two of which can be applied to all four PSP and three only to the Response Chain pattern. Thus, we cannot generalize our catalog to other patterns collected in~\cite{AutiliGLPT15}.

\section{Related Work}\label{sec:related}

Many approaches have been proposed to provide evidence about the correct behavior of SAS in face of uncertainties, particularly contributions for quantitative runtime verification~\cite{QoSMoS,calinescu2017engineering,CailliauL19:TAAS, filieri2011run,Solano+2019}.
QoSMoS~\cite{QoSMoS} pioneered the work on quantitative runtime verification of service-based systems to reach quality-of-service requirements by dynamically adapting to uncertainties. ENTRUST~\cite{calinescu2017engineering} systematically engineers trustworthy SAS by combining design-time and runtime modeling and verification to build assurance cases for SAS, dynamically updated after a system reconfiguration. Caillau and Lamsweerde~\cite{CailliauL19:TAAS} proposed an obstacle-driven runtime adaptation where monitored satisfaction rates, obtained via probabilistic assertions, are defined in terms of observed states and behaviors. The goal satisfaction rate then guides adaptation strategies.
Solano et al.~\cite{Solano+2019} proposed a goal-oriented adaptation approach with runtime verification based on reliability and cost formula derived from runtime-efficient model checking principles~\cite{filieri2011run, Calinescu:TSE21} to verify and control SAS. 
However, in all those works changes in requirements require new adaptation cycles to be started.

To incrementally tame requirements uncertainty at runtime, D’Ippolito et al.~\cite{Dippolito:ICSE14} propose a multi-tier control synthesis for adaptive systems.
Li et al.~\cite{Li:RE2022} extend those principles by proposing an iterative adaptation cycle via a multi-grained requirements relaxation. Compositional means to reduce computational overhead in runtime verification have also been proposed. 
Borda et al. propose the Adaptive CSP language~\cite{Borda:SEAMS2018} to modularly model and compositionally verify self-adaptive cyber-physical systems. 
While these approaches provide more flexibility for requirements changes, they are still limited by not supporting adaptations of individual properties. 

Changing requirements may also comprise evolving requirements. Evolution Requirements (EvoReqs)~\cite{Souza:Evolution13} focus on requirements that cause the evolution of other requirements by executing adaptation strategies in response to failures particularly related to Awareness Requirements (AwReqs)~\cite{silva2011awareness}.
Similar to the FLAGS approach~\cite{baresi2010fuzzy} the granularity of their changes goes into the goal level (a more coarse-grained way) and either delegates the changes to the target system (EvoReq) or deals with change via fuzzy levels of adaptation goals (FLAGS). Whittle et al. presented RELAX~\cite{whittle2010relax}, a requirements specification language specifically suited for SAS. Like FLAGS, they also explore fuzzy logic, but particularly to express system properties that may be relaxed due to uncertainty. 
In~\cite{Inverardi:Runtime2011}, the authors propose an approach to make requirements consistent with their corresponding evolution. Our approach differs from all those works as it focuses on a fine-grained perspective of requirements changes while also managing their change at runtime without requiring a new adaptation cycle to be started. Weyns and Iftikhar have recently provided an extension to ActivFORMS where their approach also takes into account evolution~\cite{activforms}. Similar to our work, they offer basic support for changing adaptation goals and updating the verified models of the feedback loop on-the-fly to meet evolving goals. 
However, we differ by exploring knowledge reuse in the runtime verification while dynamically adapting the property that co-evolves with the PSP-based requirement. 

Finally, integrating structured grammar with runtime verification is not novel. Perez et al.~\cite{perez2022automated} propose a framework where requirements are written in structured natural language and then transformed into monitors to analyze requirements against C code. While Mallozzi et al.~\cite{Mallozzi+2019} create observers from PSP-based requirements manually, Simmonds et al.~\cite{Simmonds+2009} model properties as sequence diagrams based on templates for PSP, from which observers are generated. Despite the notable contribution of those approaches, they are not suited for seamlessly verifying changing requirements at runtime as the property adaptation cannot be performed dynamically.

\section{Conclusion}\label{sec:conclusion}
In this paper, we addressed the need of handling requirements changes in SAS in the context of runtime verification. Particularly, we proposed a runtime verification approach where requirements are formalized with the help of property specification patterns (PSP) and verified at runtime with the help of observers. These observers are generated from templates provided by our PSP catalog, which enables precise semantics for the observers and our verification technique. To support requirements changes during runtime verification, we proposed property adaptation patterns (PAP) that define adaptations of observers at the pattern level with precise semantics. These PAP are leveraged at runtime to dynamically and safely adapt concrete observers to reflect changing requirements while preserving the knowledge about the managed system and environment gathered thus far. 
We demonstrated empirically the efficiency and correctness of our runtime verification approach for SAS with experiments and Timescales. Moreover, we discussed that continuous, incremental verification under changing properties improves the trustworthiness of the BSN.

As future work, we plan to extend our PSP/PAP catalog for runtime verification to cover all specification patterns of the catalogs for design-time model checking~\cite{AutiliGLPT15,vogel2023property} concerning safety and time-bounded liveness properties that can be verified at runtime, and develop PAP for them. Furthermore, we want to support probabilistic properties that are often used to formalize uncertain requirements in SAS. 

\clearpage
\balance
\bibliographystyle{IEEEtran}
\bibliography{bibliography}
\end{document}